\documentclass[sigconf]{acmart}

\usepackage{booktabs} 
\usepackage{subcaption} 
\usepackage{adjustbox}
\usepackage[ruled]{algorithm2e}
\setcopyright{rightsretained}

\acmDOI{10.475/123_4}

\acmISBN{123-4567-24-567/08/06}

\acmConference[MLG'18]{ACM KDD conference}{August 2018}{London, England UK}
\acmYear{2018}
\copyrightyear{2018}

\acmArticle{4}
\acmPrice{15.00}

\begin{document}
\title{
Temporal Analysis of Reddit Networks via Role Embeddings 
}

\author{Siobh\'{a}n Grayson}
\orcid{1234-5678-9012}
\affiliation{%
  \institution{Insight Centre for Data Analytics}
  \city{Dublin}
  \country{Ireland}
}
\email{siobhan.grayson@insight-centre.org}

\author{Derek Greene}
\affiliation{%
  \institution{Insight Centre for Data Analytics}
  \city{Dublin}
  \country{Ireland}
}
\email{derek.greene@insight-centre.org}

\renewcommand{\shortauthors}{S. Grayson et al.}

\begin{abstract}
Inspired by diachronic word analysis from the field of natural
language processing, we propose an approach for uncovering temporal insights regarding user roles from social networks using graph embedding methods. Specifically, we apply the role embedding algorithm, \textit{struc2vec}, to a collection of social networks exhibiting either ``loyal'' or ``vagrant'' characteristics derived from the popular online social news aggregation website Reddit. For each subreddit, we extract nine months of data and create network role embeddings on consecutive time windows. We are then able to compare and contrast how user roles change over time by aligning the resulting temporal embeddings spaces. In particular, we analyse temporal role embeddings
from an individual and a community-level perspective for both loyal and vagrant communities present on Reddit.

\end{abstract}

%
%



\maketitle

\section{Introduction}

Embeddings are now a common component of the typical text analysis pipeline thanks to their ability to apply vectorially the logic of ``You shall know a word by the company it keeps'' (Firth, J. R. 1957) and the accessibility of Mikolov's \textit{word2vec} \cite{mikolov2013distributed}. Not only have word embeddings enhanced translation tasks \cite{zou2013bilingual} but they have also exposed the cultural biases that are emeshed within languages \cite{bolukbasi2016man, grayson2017exploring}. Embeddings have even been extended to study diachronic language characteristics \cite{Kulkarni2015Statistically}. Diachronic embeddings are created by embedding separate time windows and then aligning the resulting spaces orthogonally such that distances are not warped. This allows for direct measurements to be taken of how much the meaning of semantically similar words change over time in comparison to eachother using distance metrics such as cosine distance \cite{hamilton2016diachronic}. 

Therefore, motivated by the concepts and findings being developed for diachronic word embeddings, in this paper we explore how the application of the same principles can be leveraged to study structural roles from a temporal perspective. In the same way words with a similar meaning will repeatability appear in the same contexts, structural roles in graphs are also defined by the topological company that they keep. However, structurally equivalent roles may or may not occur in close proximity within a graph \cite{Donnat2018Learning}. Hence, by embedding networks into different dimensions the distances between similar entities can be reduced. Our goal is to then map the participants of the popular social media website \textit{Reddit}\footnote{The url address for this site is: https://www.reddit.com/}, into an embedding space that best represents the structural roles that they occupy and to then measure how their roles change over time. In particular, we analysis how roles evolve from both an individual and community level perspective. Our findings suggest that while participant roles fluctuate a lot, the ubiquitous community roles present are relatively static in comparison.

\begin{figure*}[h!]

\begin{subfigure}[b]{0.24\textwidth}
	\includegraphics[width=\linewidth]{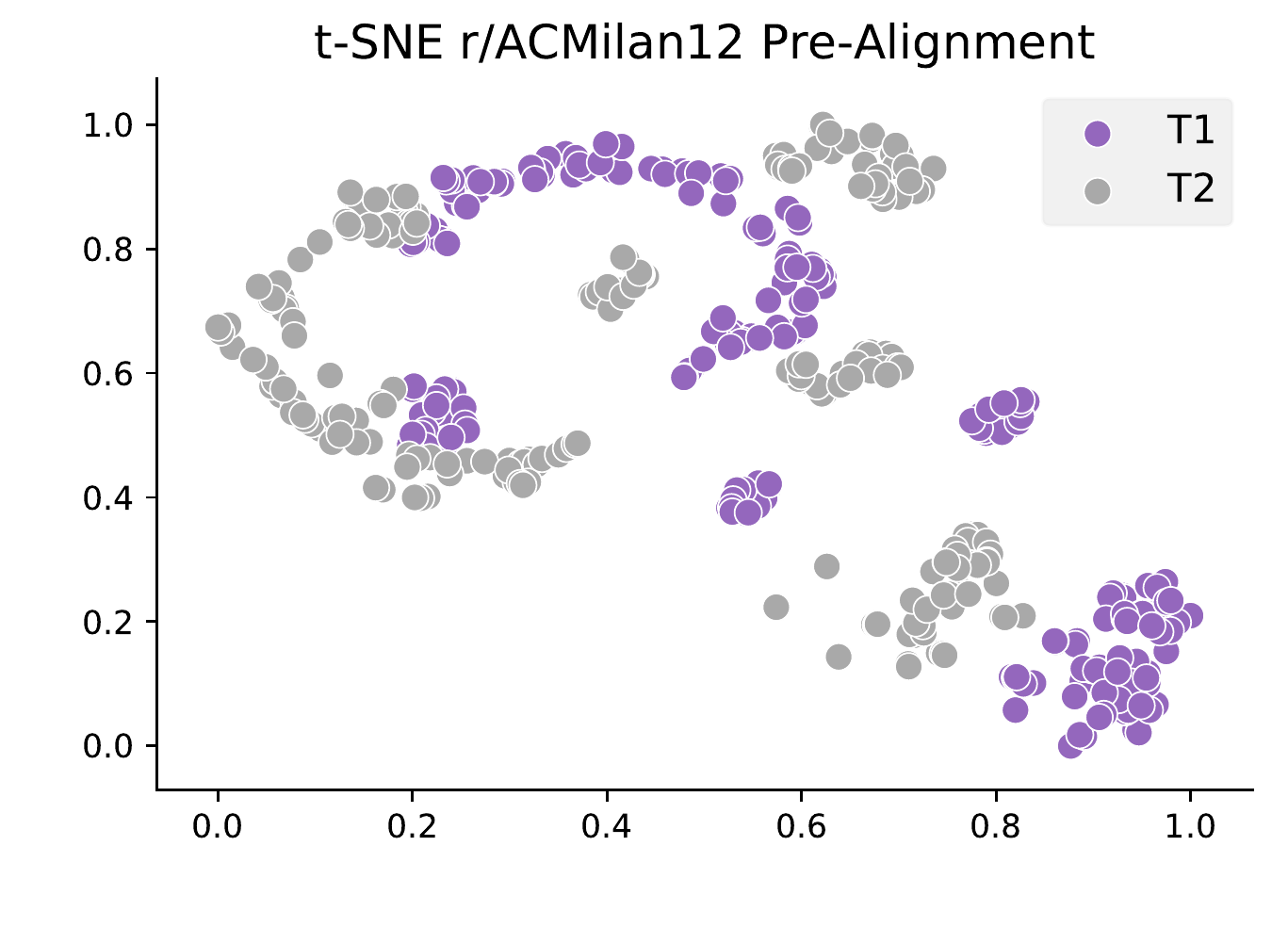}
	\caption{Pre-Alignment Embeddings}
	\label{fig:}
\end{subfigure}
\begin{subfigure}[b]{0.24\textwidth}
	\includegraphics[width=\linewidth]{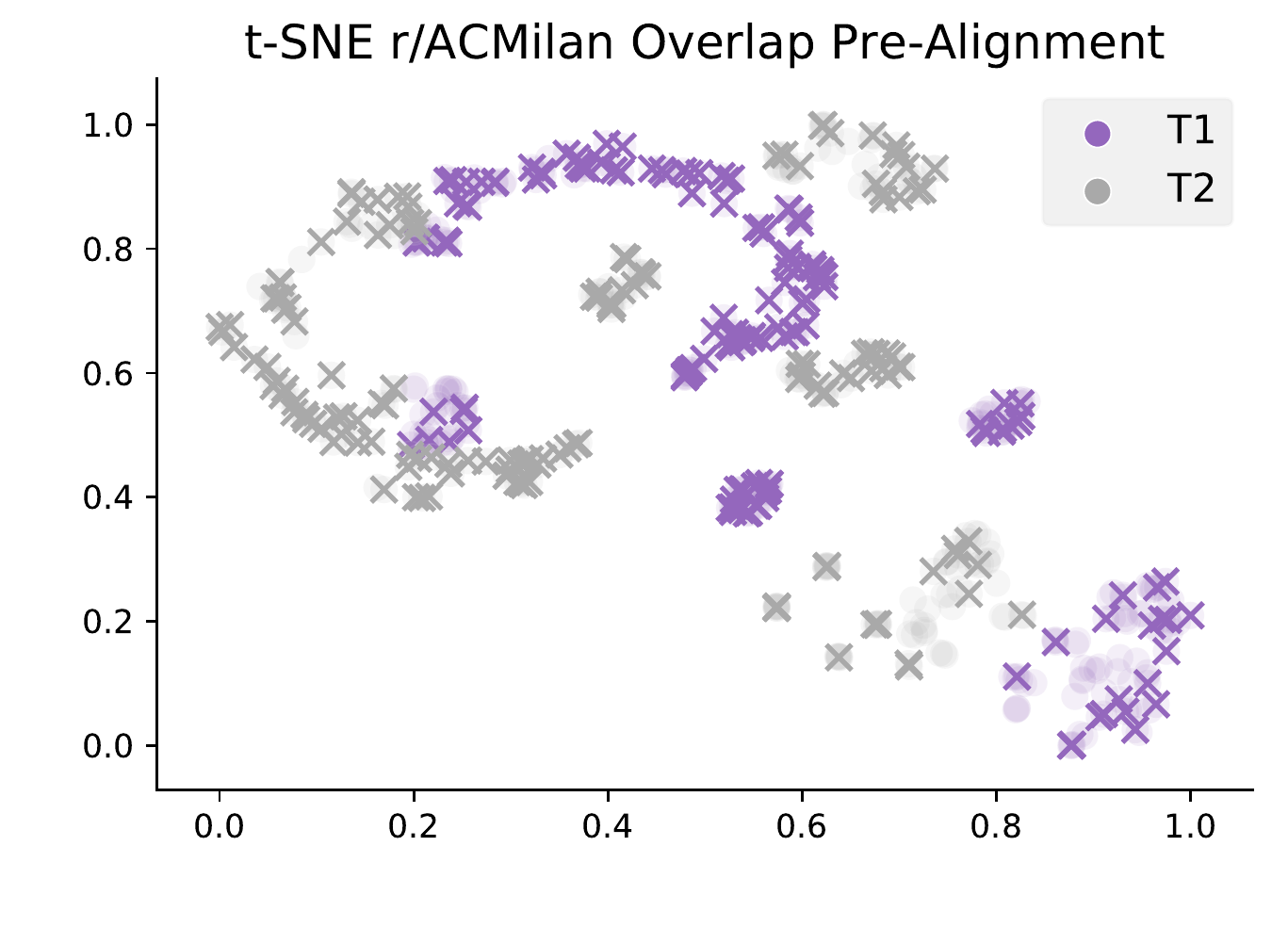}
	\caption{User Overlap Embeddings}
	\label{fig:gull2}
\end{subfigure}
\begin{subfigure}[b]{0.24\textwidth}
	\includegraphics[width=\linewidth]{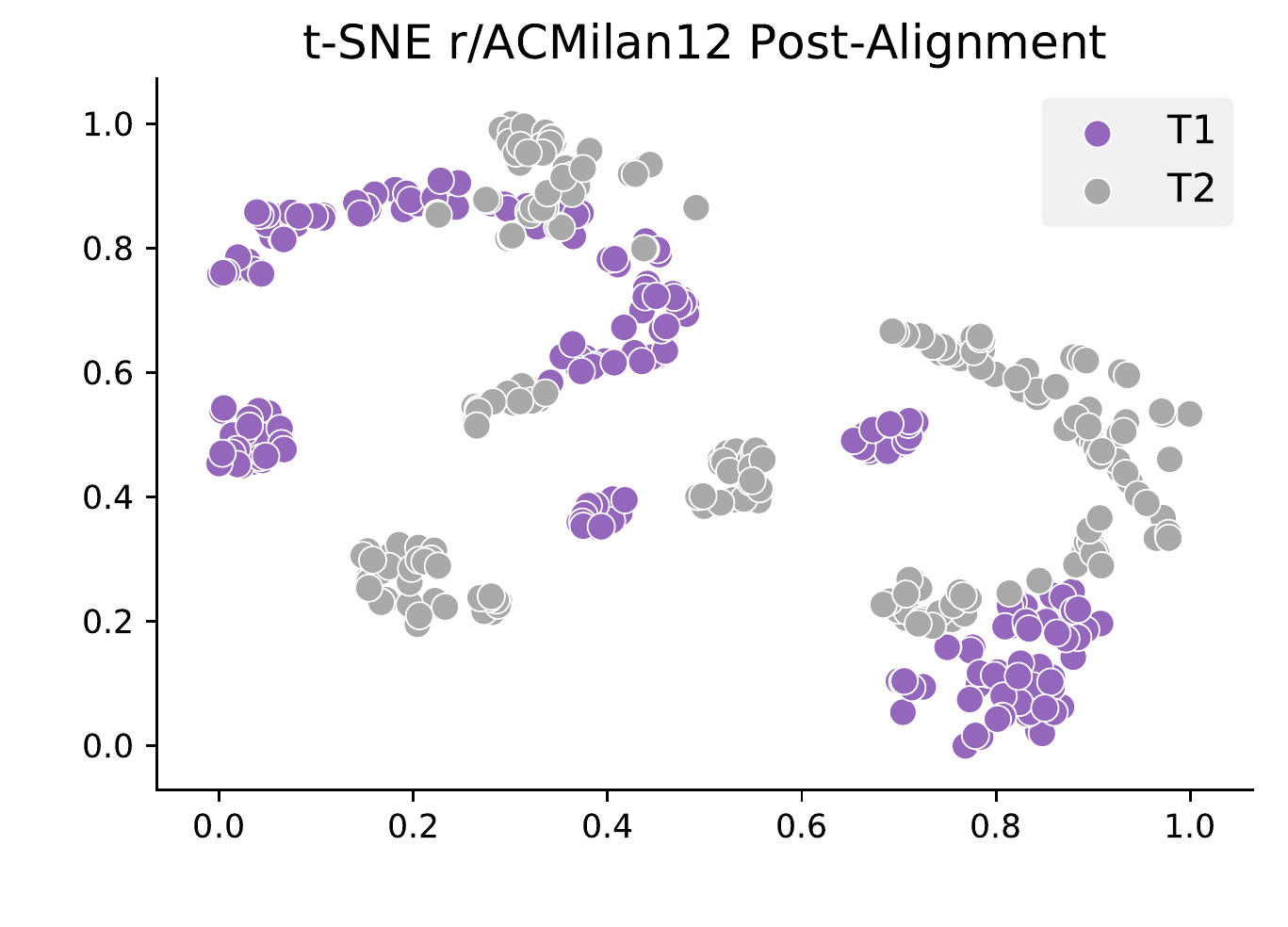}
	\caption{Post-Alignment Embeddings}
	\label{fig:gull2}
\end{subfigure}
\begin{subfigure}[b]{0.24\textwidth}
	\includegraphics[width=\linewidth]{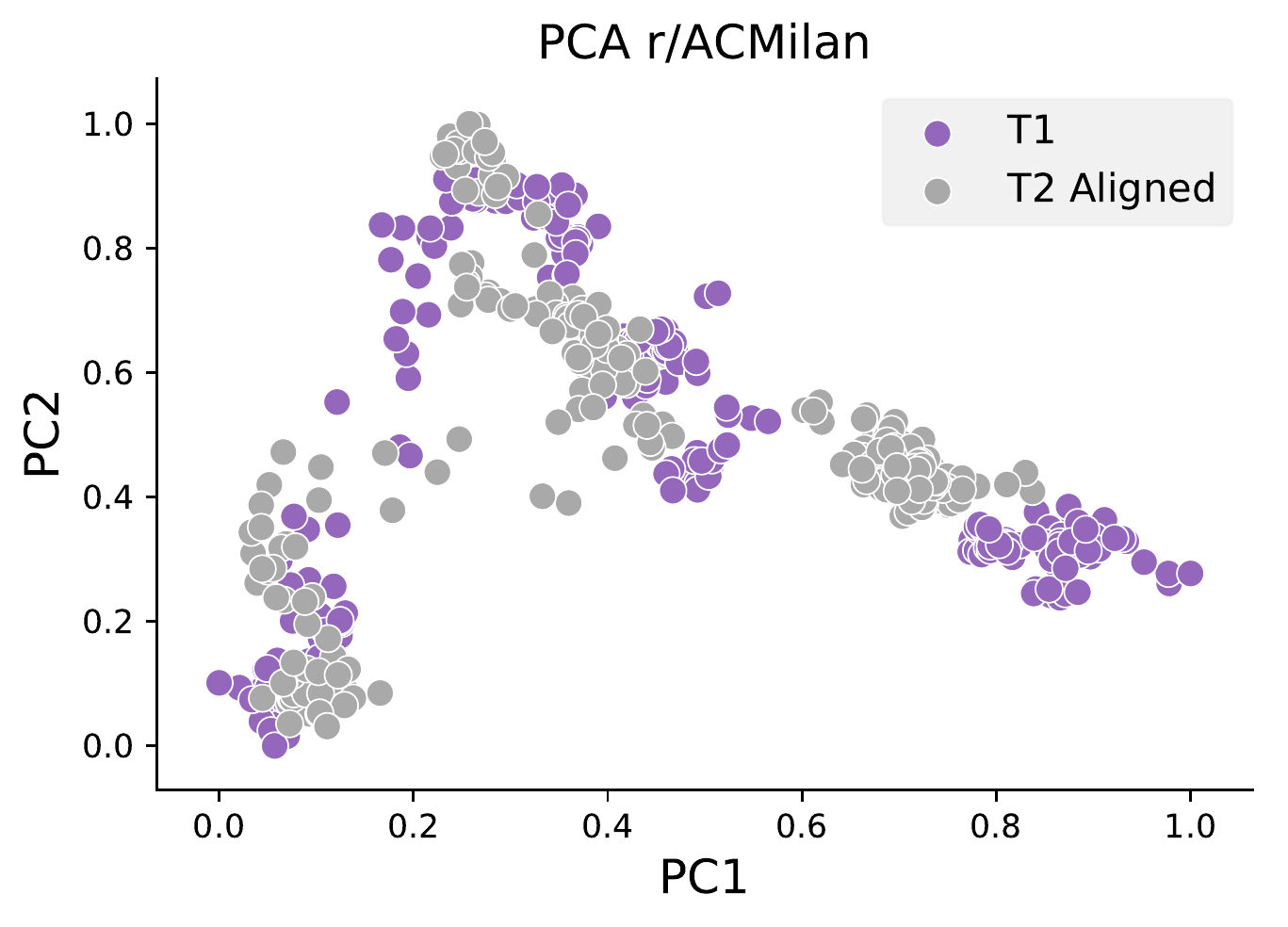}
	\caption{PCA Embeddings}	
\end{subfigure}
\caption{Visualisation of the loyal subreddit `r/ACMillan' before and after alignment and dimension reduction. }
\label{fig:ACMilan}
\end{figure*}

\section{Related Work}

The social news aggregation website Reddit was founded in 2005 and has grown over the years to now have over 200 million unique users. Fives years of Reddit's lifetime has been analysed by Singer et al. \cite{Singer2014Evolution} from both a user submissions perspective, to how community level attention evolves over time, resulting in ``an ever-increasing diversification of topics accompanied by a simultaneous concentration towards a few selected domains''. In 2016, Newell et al. conducted a study on user migration in social networks and found that an important factor in Reddit's ability to retain users was the availability of ``niche'' content not provided elsewhere \cite{Newell2016User}. Thus, this retention of users suggests a certain level of loyal activity which is explored in further detail by Hamilton and Zhang et al. in 2017 who categorise different subreddits as either ``loyal'' or ``vagrant'' communities based on the network and textual characteristics \cite{Hamilton2017Loyalty}. The vagrant nature of Reddit is also investigated by Leavitt in 2015 via ``throwaway accounts'' \cite{Leavitt2015This}, potentially one of the shortest temporal roles present on Reddit. 


\subsection{Graph Embeddings}

Increasingly, network scientists are adpoting embedding techniques of their own to examine graphs in different dimensions. Grover et al. even gestering a nod to \textit{word2vec} with their graph embedding algorithm entitled \textit{node2vec} \cite{grover2016node2vec}. In short, a diverse range of embedding approaches exist ranging from granular node and edge generated emebeddings  \cite{wang2017community, cavallari2017learning} to whole graph \cite{niepert2016learning} and dynamic \cite{Zhou2018Dynamic}. In fact, there exists at least three comprehensive survey papers on the subject \cite{wang2017knowledge, nishana2013graph, cai2018comprehensive}. Reddit is even amoung one of the datasets used by Hamilton et al. to evaluate their attributed graph embedding algorithm \textit{GraphSAGE} \cite{hamilton2017inductive}. As such, network scientists have a large pool of techniques to select from, each catering for different types of graphs. Although dynamic behaviour has been modelled before \cite{Rossi2013Modeling}, dynamic embedding algorithms specifically designed for finding structural equivalences that also incorporate direction and weight are few and far between. For the purposes of this study, three role embedding techniques stood out, Rolx \cite{Henderson2012RolX}, GraphWave \cite{Donnat2018Learning}, and struc2vec \cite{Ribeiro2017Struc2Vec}. After testing each, we decided to remain with struc2vec as it could be optimised such that computations completed faster than it peers and it generalised well, grouping similar roles together as opposed to over-fitting and identifying them as completely different to each other.

%
%
%
%
%

\section{Methodology}

In parallel to our motivation to observe whether temporal role variations occur at user and subreddit levels, we're also keen to learn whether role variation is also related to subreddit type. Therefore, we're using a subset of the directed Reddit chain-based interaction networks where users are linked if they comment within a linear chain originally curated by Hamilton and Zhangs in their work on characterising Reddit loyalty\cite{Hamilton2017Loyalty}\footnote{Further details can be found on the webpage where the dataset is available to download: http://snap.stanford.edu/data/web-RedditNetworks.html}. Our dataset consists of 16 subreddits identified by Hamilton et al. as  exhibiting the most ``loyal'' user features (teams and sports related subreddits) and 13 subreddits identified as having the highest ``vagrant'' user patterns. When identifying loyal and vagrant communities, Hamilton et al.  considered user commenting behaviour on Reddit over time. They define loyal and vagrant users as follows:
\begin{itemize}
	\item Loyal members are users who for two consecutive months have submitted at least 50\% of their comments to one Subreddit. In doing so, they exhibit a preference and commitment to this Subreddit.
	\item Vagrant members on the other hand are defined as users  who comment 1 to 3 times within a Subreddit in one month but then do not submit any comments the subsequent month despite still being active on Reddit. 
\end{itemize}
For temporal analysis, we partitioned 9 consecutive months of data, spanning from late January to October in 2014, into three temporal windows consisting each consisting of three months. A summary of this data is provided in Table \ref{tab1}. 

\begin{table}[htbp]
\caption{Summary of Reddit Data}
\centering
\begin{adjustbox}{max width=1\linewidth}
\begin{tabular}{l l rr  rr  rr}
\toprule
\textbf{Class} &
\textbf{\# SR} &
\textbf{\# V$_{T1}$} &
\textbf{\# E$_{T1}$} &
\textbf{\# V$_{T2}$} &
\textbf{\# E$_{T2}$} &
\textbf{\# V$_{T3}$} &
\textbf{\# E$_{T3}$} \\
\midrule
Loyal & 13 & 15,319 & 89,496 & 15,193 & 91,138 & 14,531  & 87,149 \\
Vagrant & 16 & 13,462 & 22,323 & 14,030 & 23,831 & 13,314 & 22,247\\
\bottomrule
\multicolumn{8}{l}{Notation - SR: Subreddits, V$_{T1}$: Nodes in Temporal Window 1,}\\
\multicolumn{8}{l}{E$_{T1}$: Edges in temporal window 1.}
\end{tabular}
\end{adjustbox}
\label{tab1}
\vspace*{-0.4cm}
\end{table}

For the purposes of this study, two users are defined as having corresponding roles if their occurrences within the Reddit networks are structurally equivalent. To assess user role variation over time, we first select the 100 highest frequency participants for each 3 months and then use the overlap of this set that spans all window partitions to extract temporally related networks. Once we have our temporal networks, actors are then described in terms of their roles by applying the directed and weighted version of the graph embedding algorithm, \textit{struc2vec} \cite{Ribeiro2017Struc2Vec}, specifically designed to capture structural equivalence between nodes. Specifically, similar roles are mapped closer together in the resulting embedding space while dissimilar roles will be further a part. Hence, role similarity can be assessed by measuring the distance between role embeddings. However, before embeddings generated from different time windows can be compared they must first be aligned.

\begin{figure*}[h!]

\begin{subfigure}[b]{0.24\textwidth}
	\includegraphics[width=\linewidth]{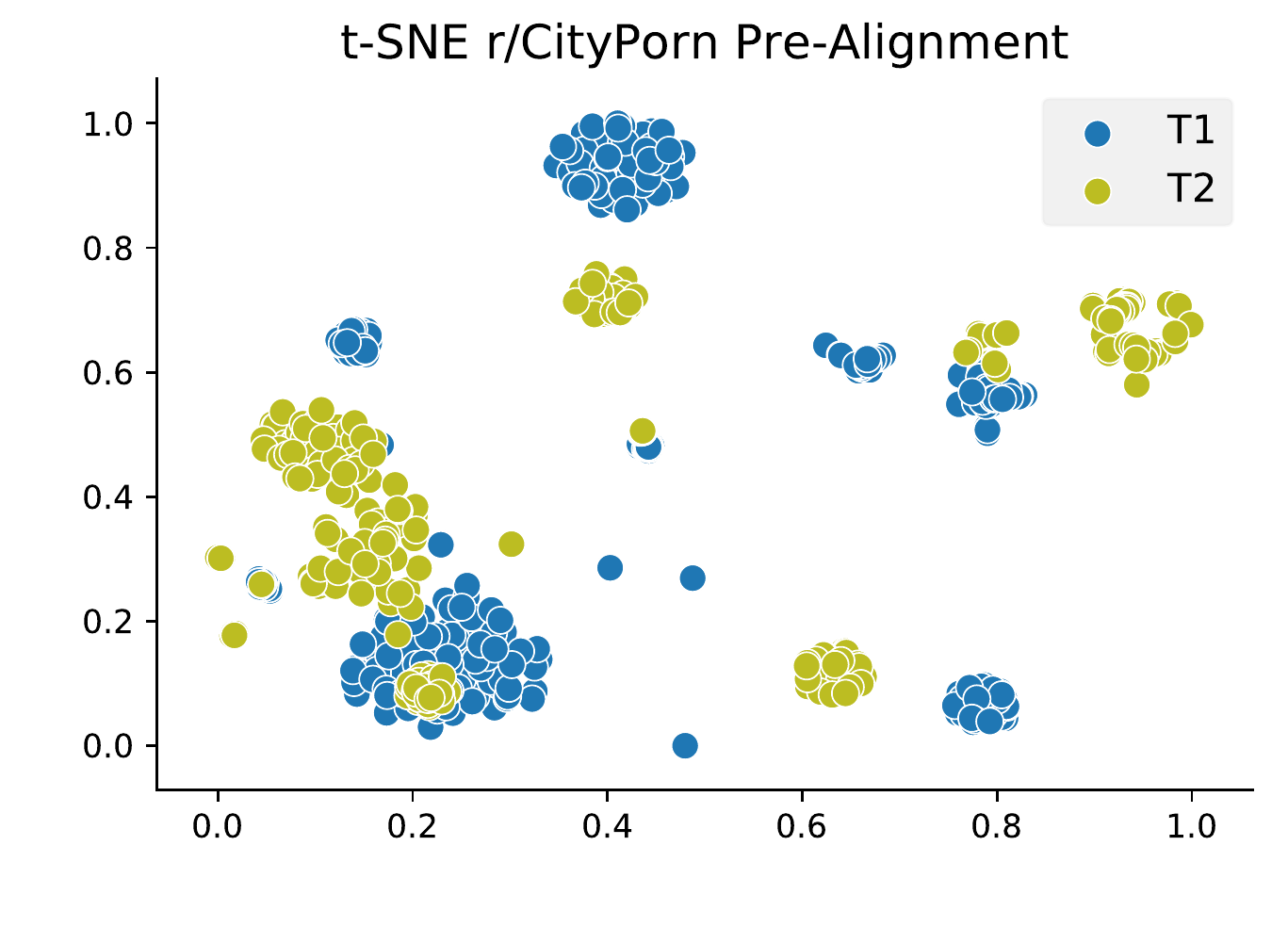}
	\caption{Pre-Alignment Embeddings}
	\label{fig:gull}
\end{subfigure}
\begin{subfigure}[b]{0.24\textwidth}
	\includegraphics[width=\linewidth]{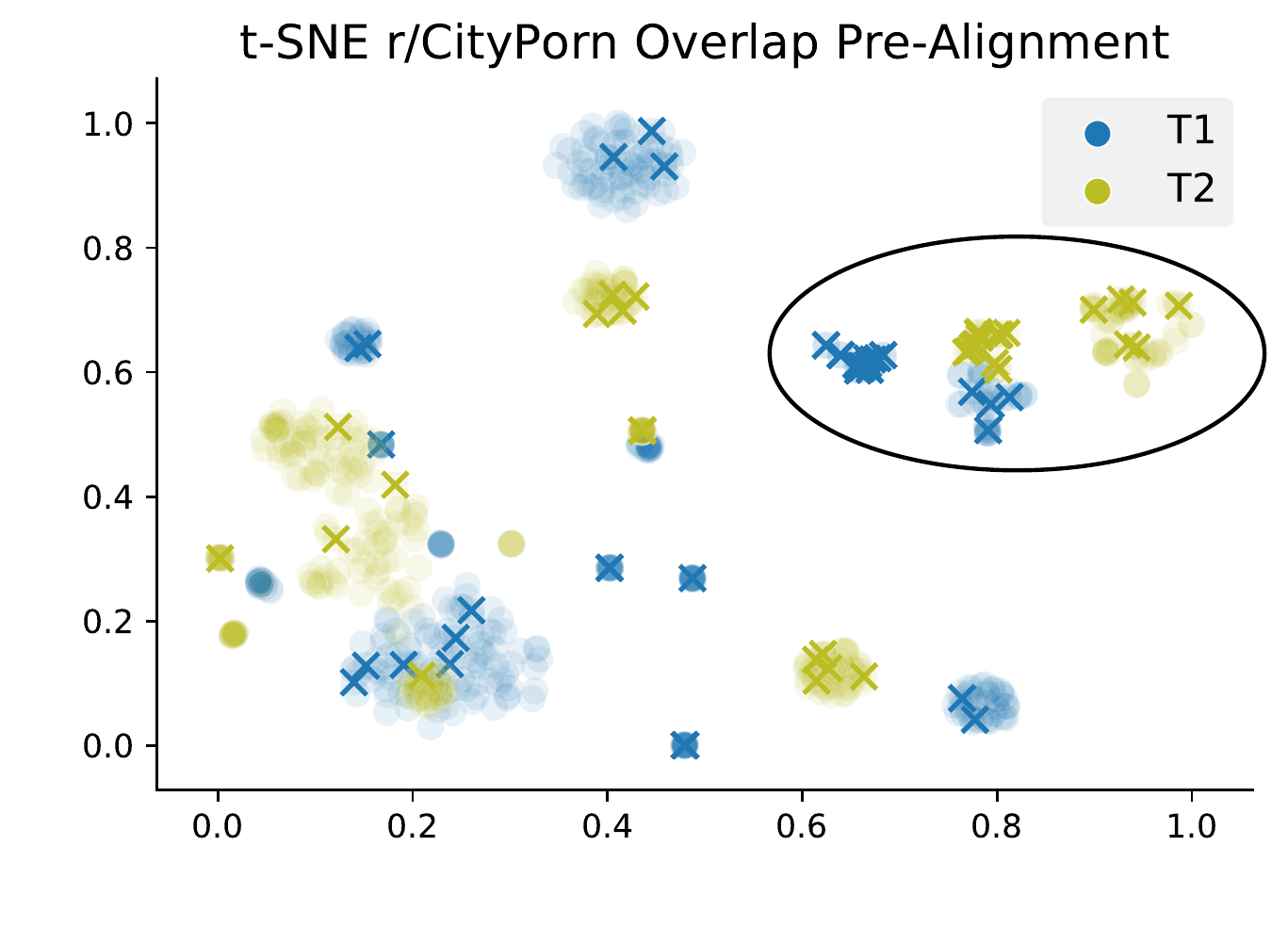}
	\caption{User Overlap Embeddings}
	\label{fig:gull2}
\end{subfigure}
\begin{subfigure}[b]{0.24\textwidth}
	\includegraphics[width=\linewidth]{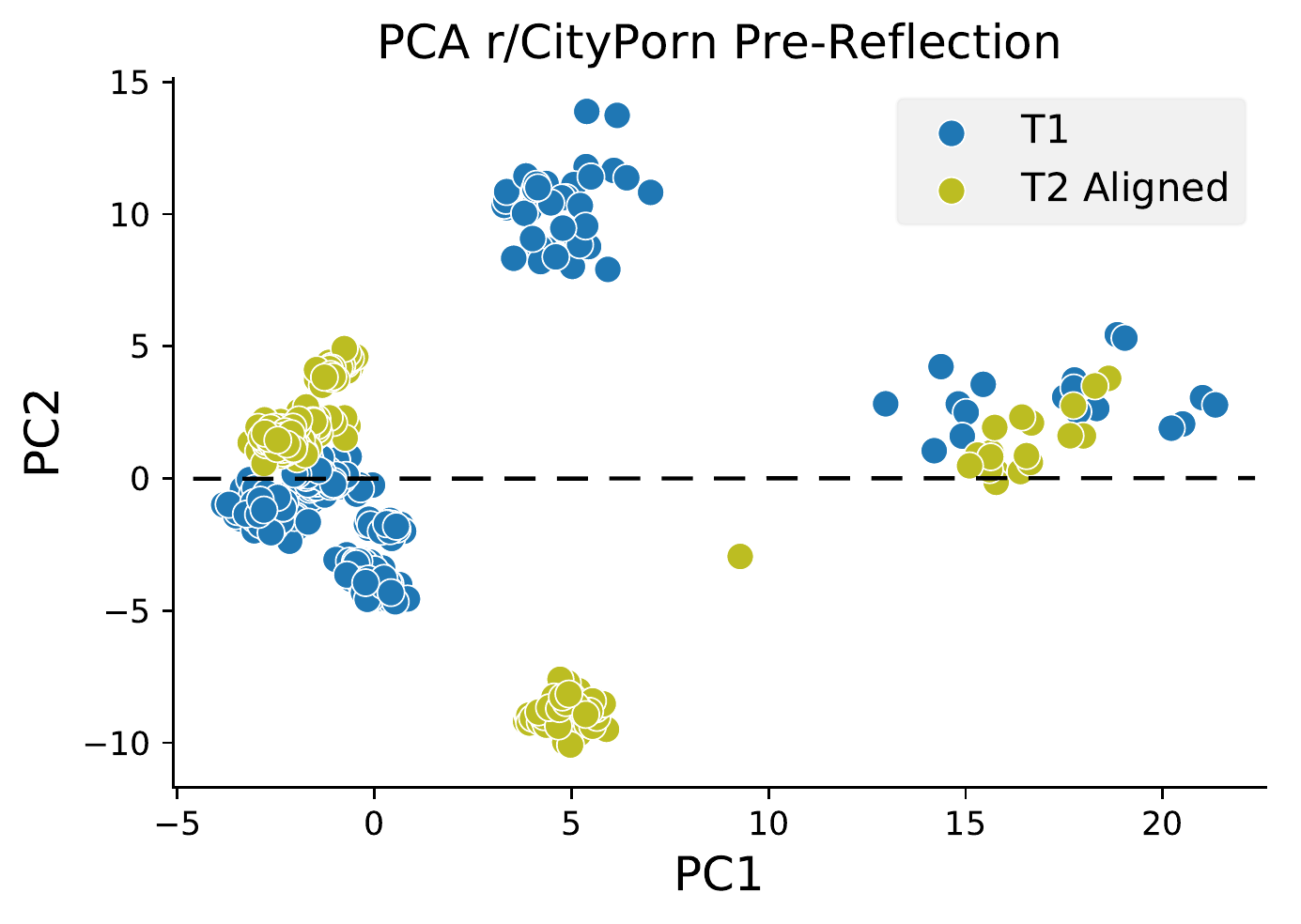}
	\caption{PCA Pre-Alignment}
	\label{fig:gull2}
\end{subfigure}
\begin{subfigure}[b]{0.24\textwidth}
	\includegraphics[width=\linewidth]{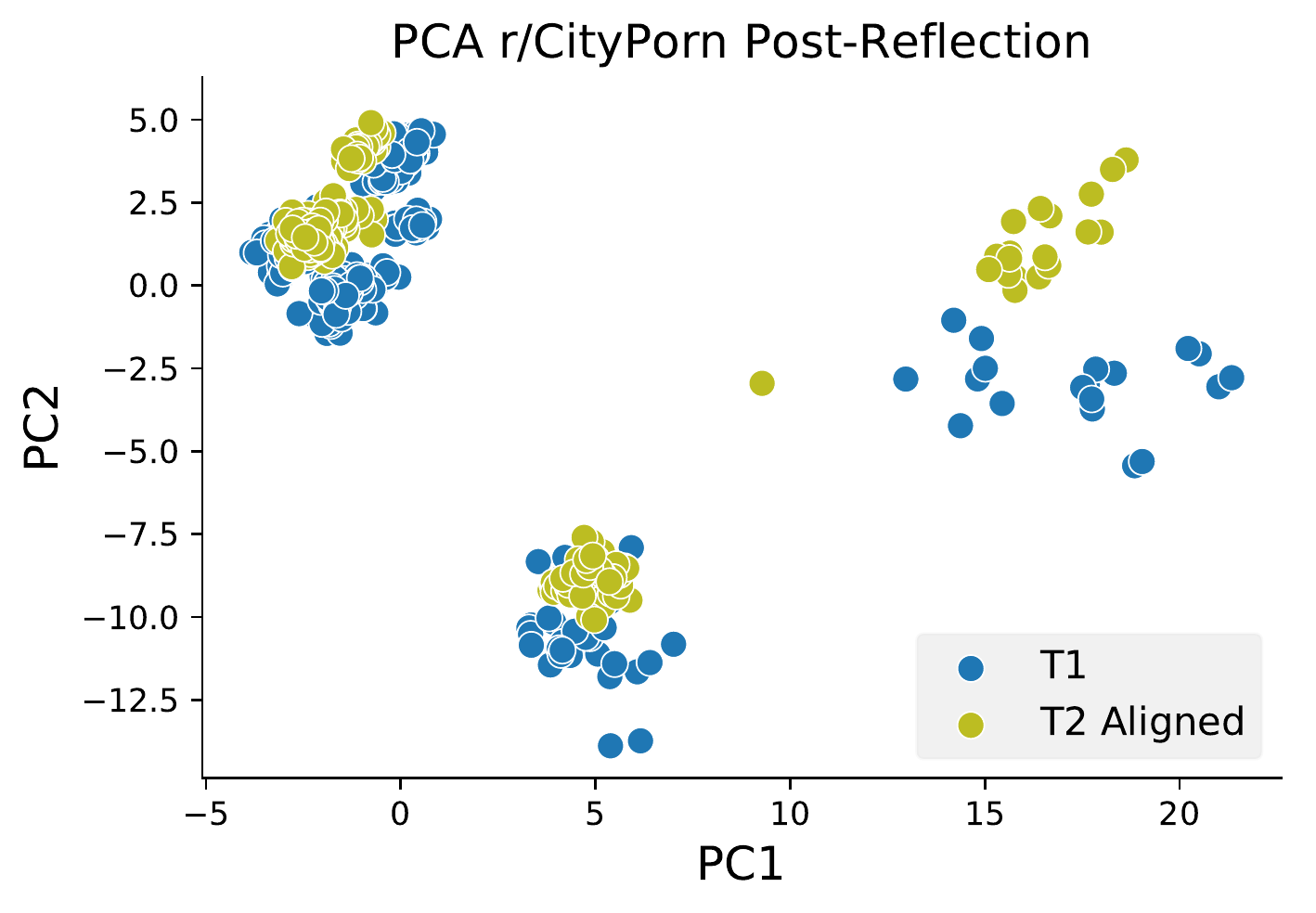}
	\caption{PCA Post-Alignment}
	\label{fig:gull2}
\end{subfigure}
\caption{Visualisation of the vagrant subreddit `r/CityPorn' before and after alignment and dimension reduction.}
\label{fig:badAlignment}
\end{figure*}

\subsection{Temporal Role Alignment}

\begin{figure}[ht]
\centerline{
\includegraphics[width=0.75\linewidth]{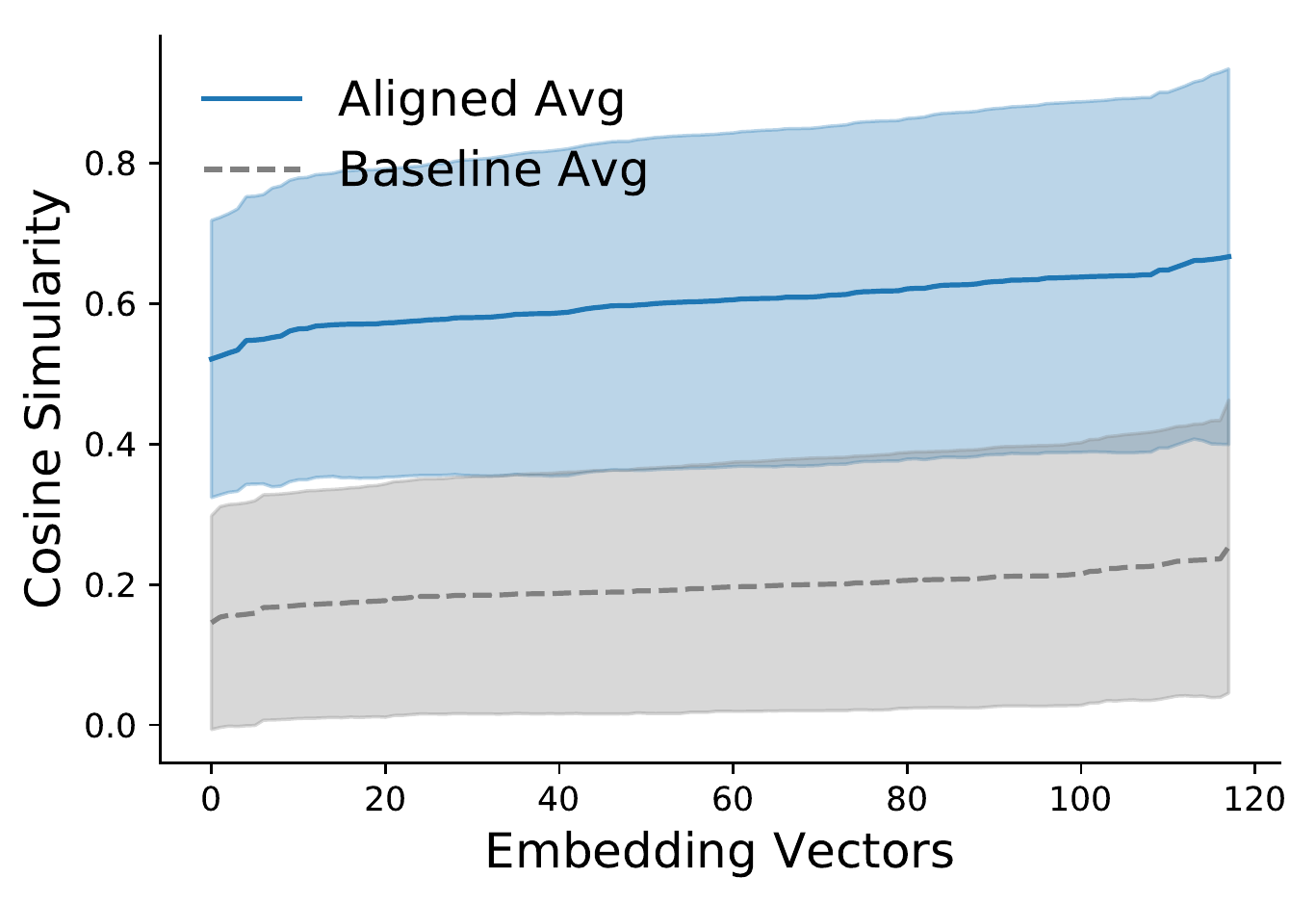}}
\caption{Cosine Similarity results for alignment evaluation.}
\label{fig:cosineSimilarity}
\end{figure}

The embedding spaces in this study are aligned using normalised orthogonal Procrustes, an approach popular for aligning diachronic word embeddings \cite{smith2016offline, hamilton2016diachronic}, as it derives the optimal rotation of a ``source'' matrix with respect to a ``target'' matrix without scaling by minimising the sum of squared distances between elements. This results in the ability to directly compare temporal embedding spaces to each other using dimension appropriate distance metrics. In particular, the orthogonal Procrustes rotation between spaces is computed by mapping the overlapping sets of users to each other. Fig.\ref{fig:ACMilan} illustrates the process by visualising T2 (time period 2) embeddings being aligned to T1 (time period 1) embeddings using t-SNE \cite{maaten2008visualizing}.
Alignments can then be evaluated by generating a second embedding matrix for the same time period and comparing the cosine similarity between vectors. Fig.\ref{fig:cosineSimilarity} displays the average of aggregated cosine similarity results ($1/N \sum^{N}_{i=1} \cos(\pmb v^t_i, \pmb v^{t+\Delta}_i)$) and the standard deviations computed across all embedding spaces and their duplicates for both before (Baseline) and after alignment. In all cases, rotations reduced the dissimilarity between temporal user embeddings.  


\begin{figure*}[ht]
\begin{subfigure}[c]{0.35\linewidth}
	\includegraphics[width=\linewidth]{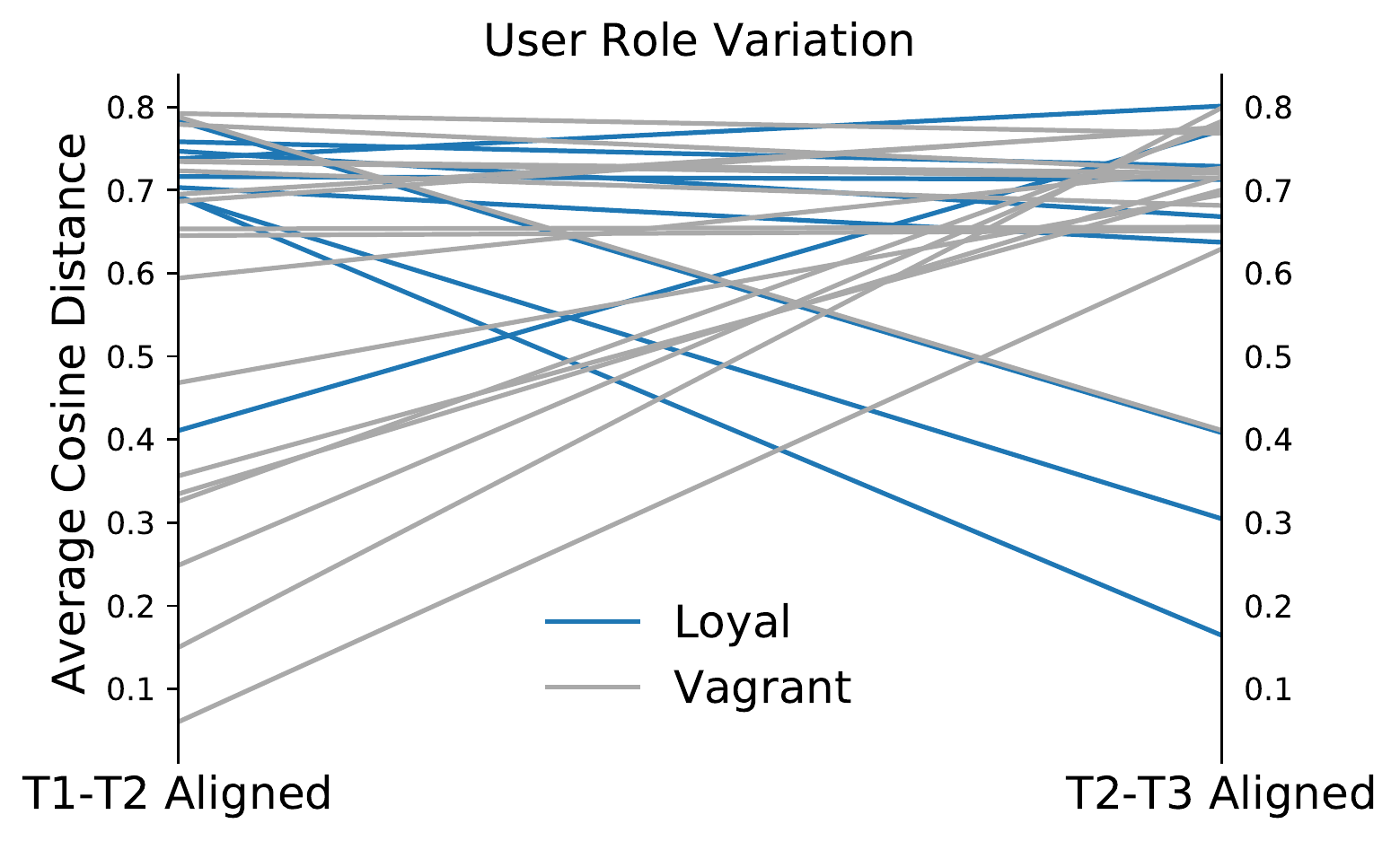}
	\caption{User roles are not static.}
	\label{fig:gull}
\end{subfigure}
\begin{subfigure}[c]{0.35\linewidth}
	\includegraphics[width=\linewidth]{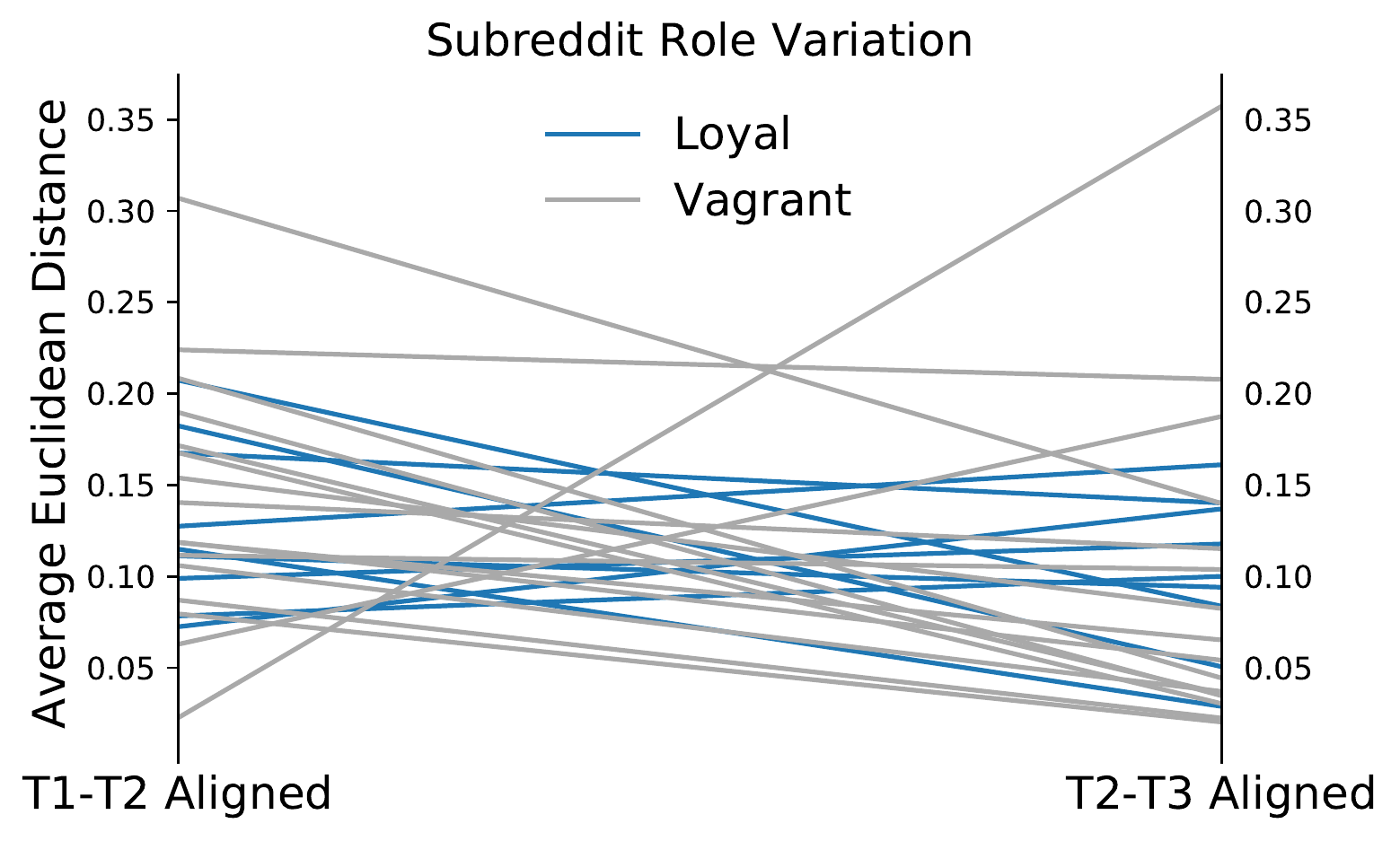}
	\caption{Community roles are relatively static.}
	\label{fig:gull2}
\end{subfigure}
\begin{subfigure}[c]{0.25\linewidth}
	\includegraphics[width=\linewidth]{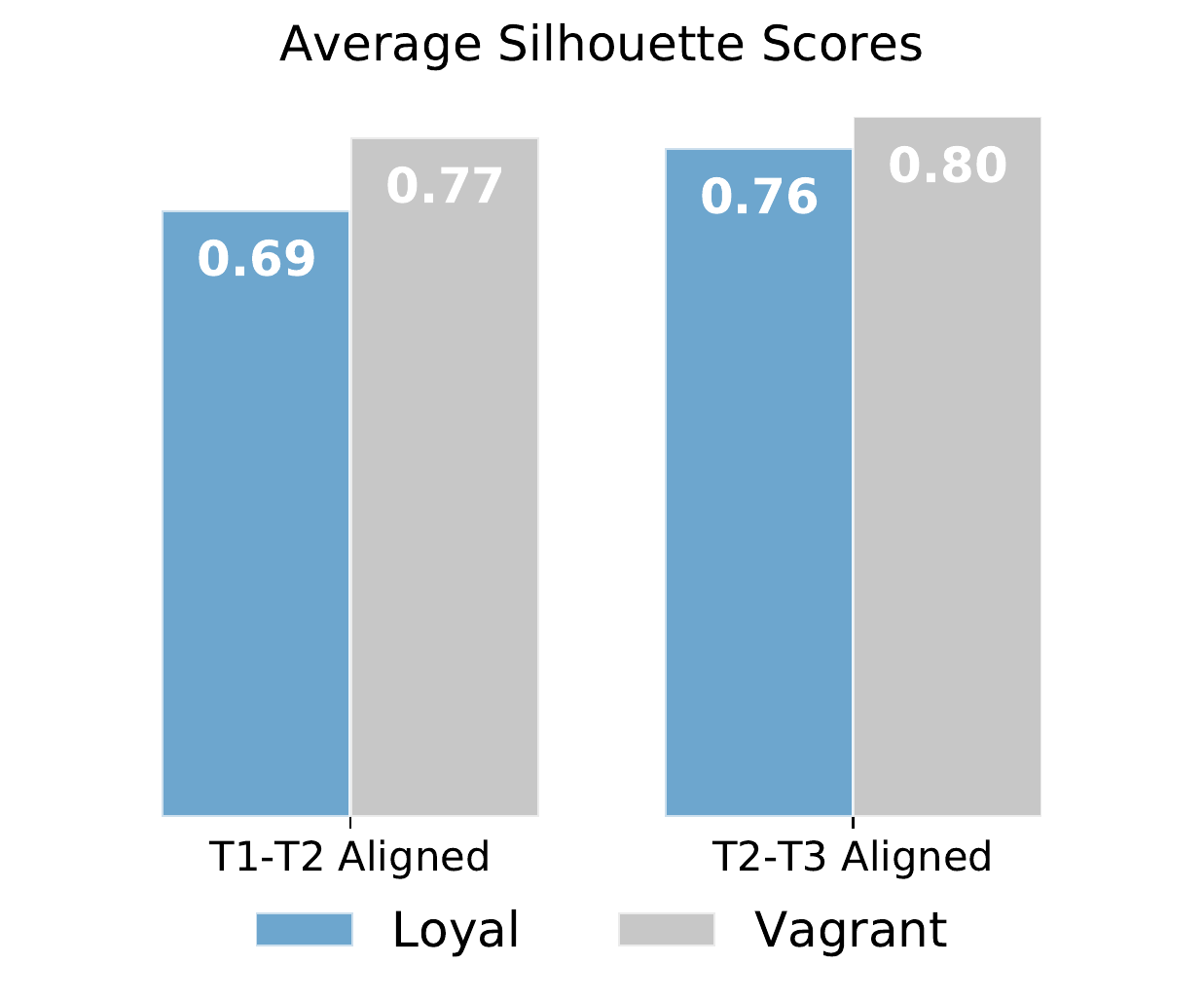}
	\caption{Variation of defined roles.}
	\label{fig:gull2}
\end{subfigure}

\caption{The temporal user and community role dynamics observed via three different metrics for comparing similarity: Cosine distance, Euclidean distance, and Silhouette scores.}
\label{fig:results}
\end{figure*}

\begin{figure}[htbp]

\begin{subfigure}[b]{0.49\linewidth}
	\includegraphics[width=\linewidth]{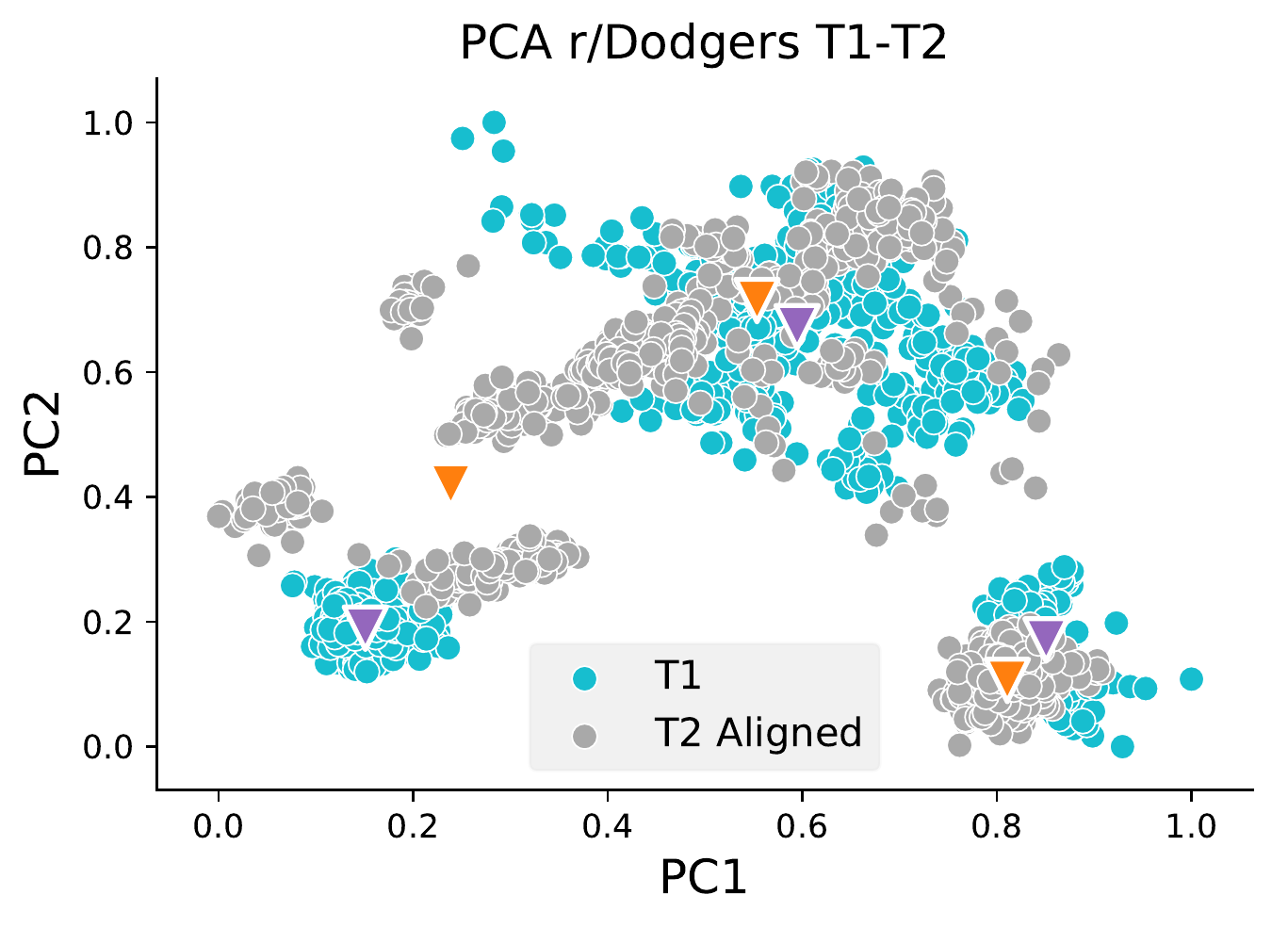}
	\caption{Loyal Subreddit}
	\label{fig:gull}
\end{subfigure}
\begin{subfigure}[b]{0.49\linewidth}
	\includegraphics[width=\linewidth]{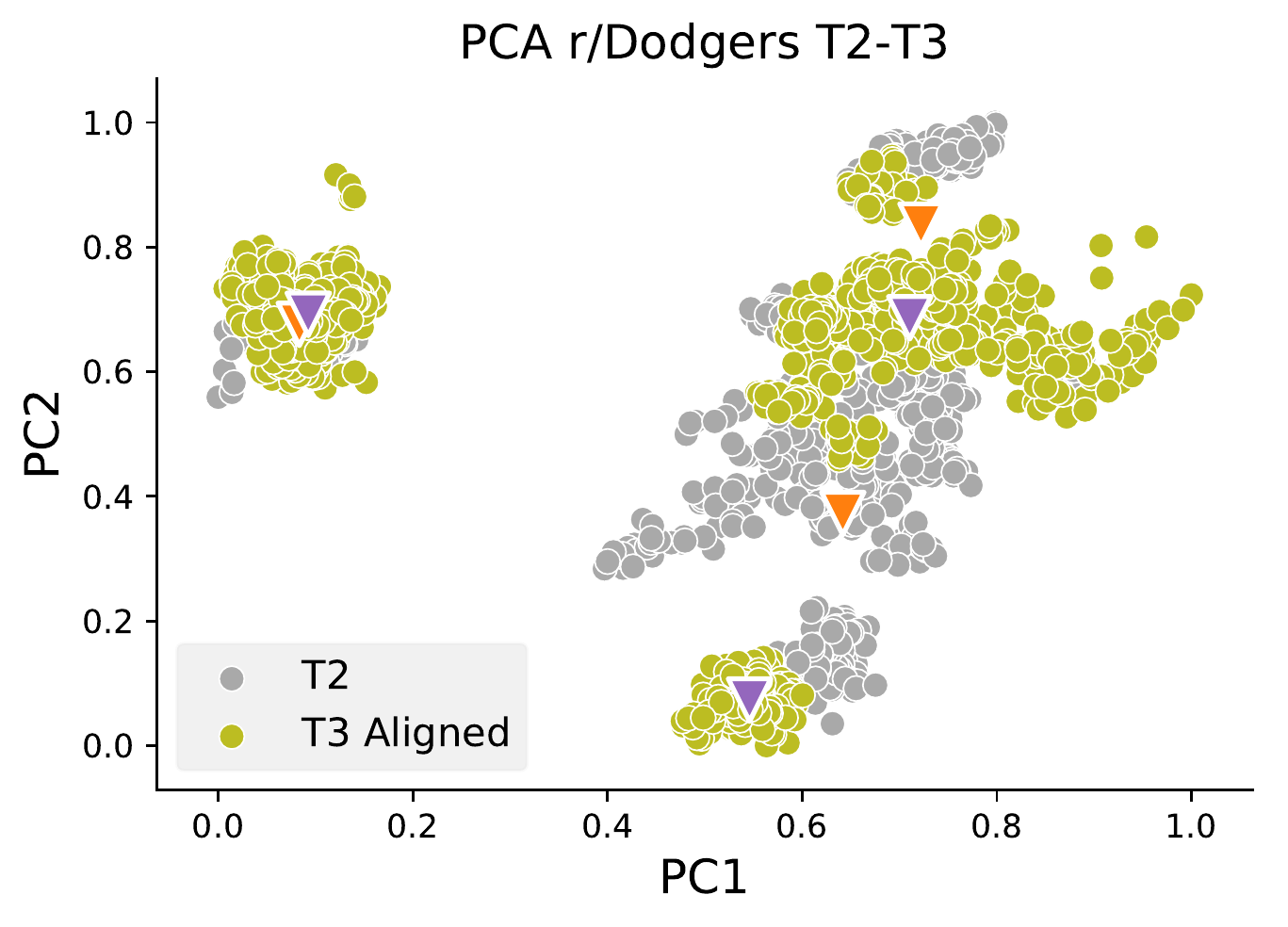}
	\caption{Loyal Subreddit}
	\label{fig:gull2}
\end{subfigure}

\begin{subfigure}[b]{0.49\linewidth}
	\includegraphics[width=\linewidth]{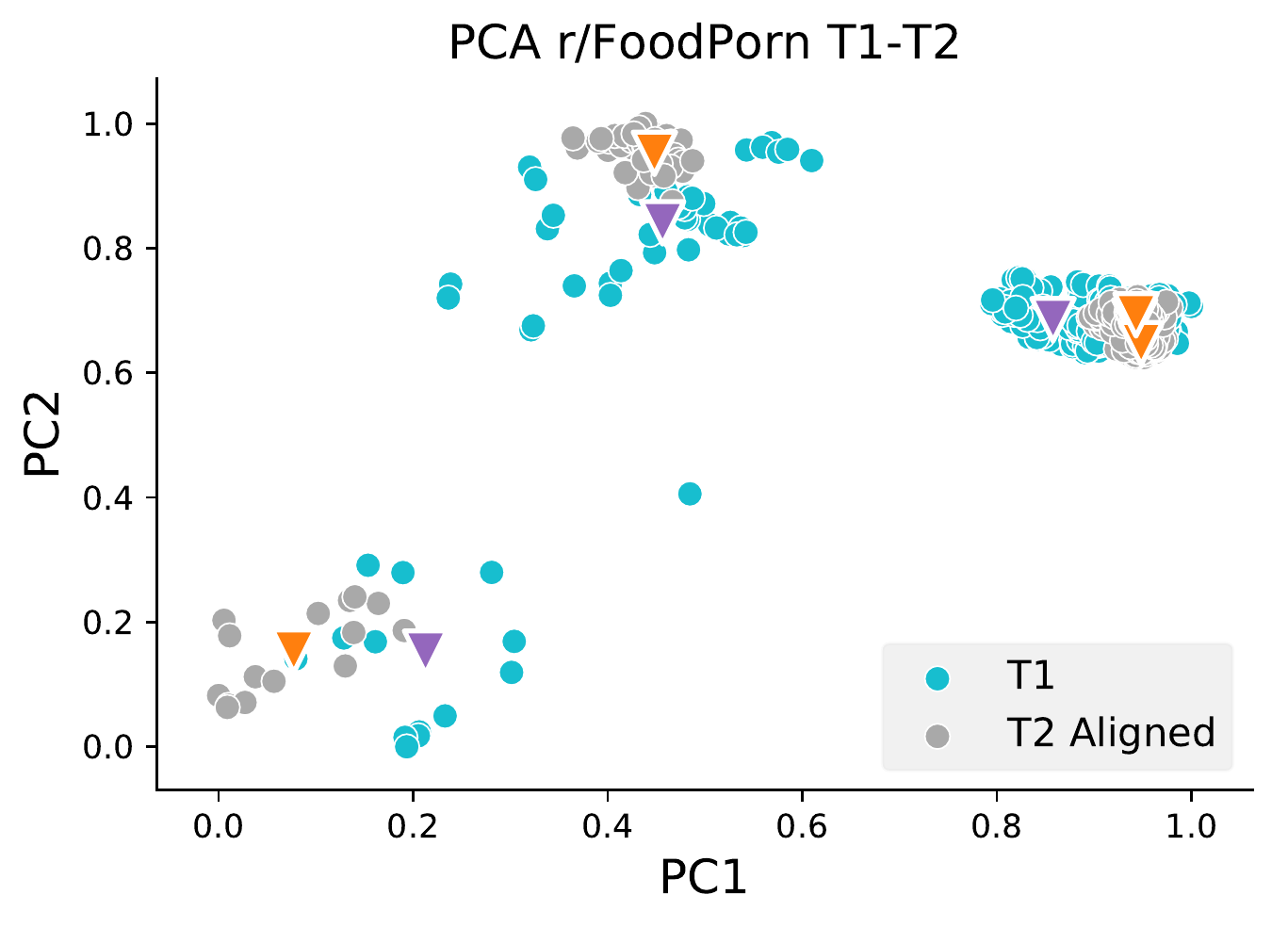}
	\caption{Vagrant Subreddit}
	\label{fig:gull2}
\end{subfigure}
\begin{subfigure}[b]{0.49\linewidth}
	\includegraphics[width=\linewidth]{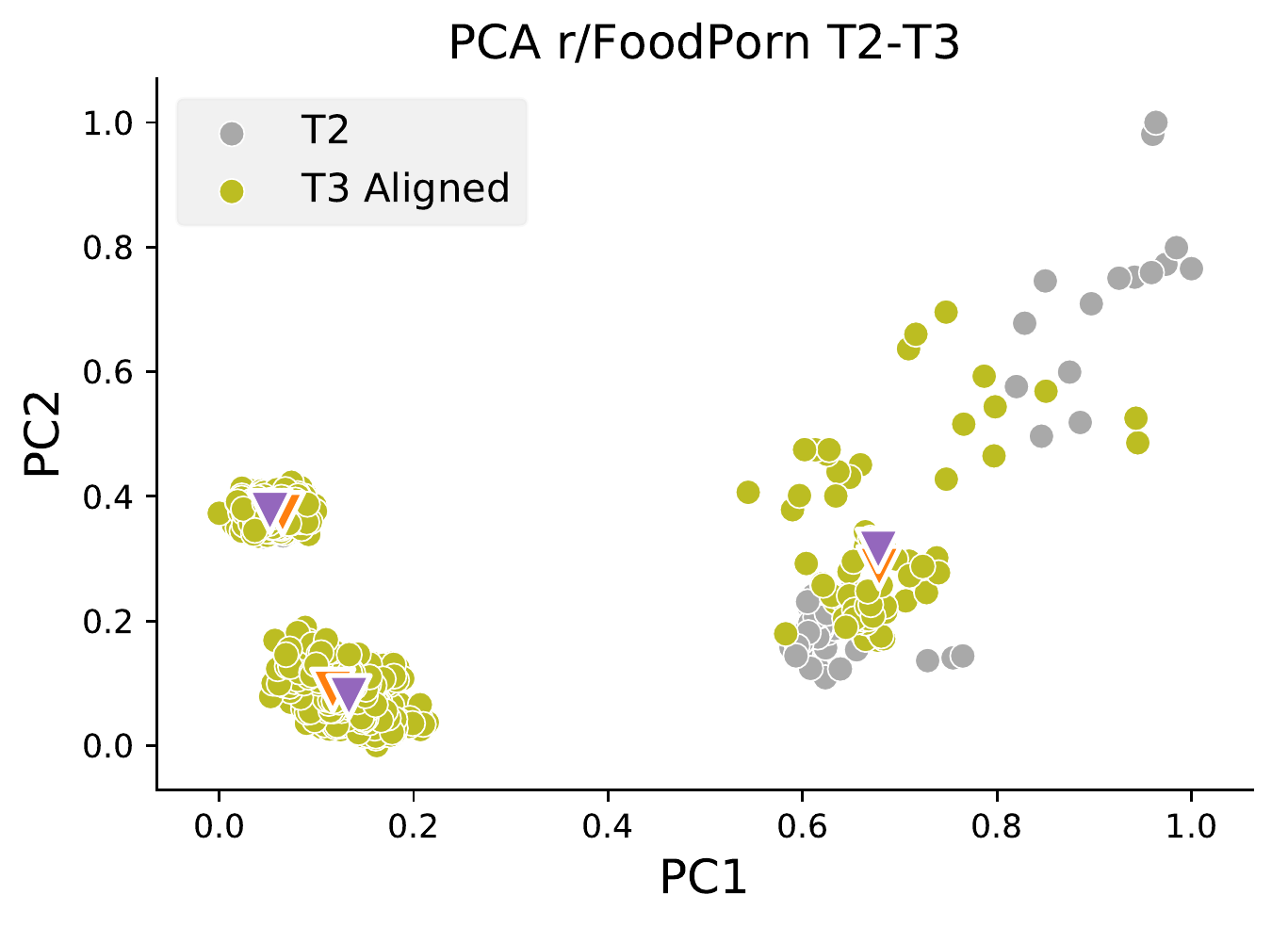}
	\caption{Vagrant Subreddit}
	\label{fig:gull2}
\end{subfigure}
\caption{Subreddit PCA projections across time.}
\label{fig:vis}
\end{figure}

However, although orthogonal Procrustes appeared to perform reasonably well when evaluated using overlapping user embeddings, it did not always correctly align community role embeddings, defined as the agglomerates of similar roles that emerge after the application of Principal Component Analysis (PCA) to aligned spaces. Specifically, the T1-T2 Aligned and T2-T3 Aligned spaces, where T1-T2 Aligned stands for the embedding space that occurs from the amalgamation of T1 with T2 embeddings that have been aligned to T1, and similarly for T2-T3 Aligned. Unlike our previous example of \textit{ACMilan} embeddings (Fig.\ref{fig:ACMilan}), the majority of overlapping individuals in \textit{CityPorn}'s embeddings are confined to a relative small region of the space (circled in Fig.\ref{fig:badAlignment}(b)). This results in the sign of eigenvectors being `flipped' during PCA and hence, rather than similar community roles overlapping, they become mirrored in the resulting PCA space (Fig.\ref{fig:badAlignment}(c). To resolve this, further alignment of roles is applied by changing the signs of equivalent principal components to agree if they do not already. 

%
%
%
%
%

\subsection{Measuring Role Variation Across Time}

Once embeddings have been aligned, temporal comparisons can be made directly using appropriate distance metrics. To detect changes in an individual's role across time, we compute the cosine distance between an actors embedding at time $t$ and $t+\Delta$: $1-\cos(\pmb v^t_i, \pmb v^{t+\Delta}_i)$. Greater distances indicate a larger deviation in the type of roles a participant occupied during different periods. While small cosine distances suggest an individual's role has not changed much over time as they map into a relatively similar space. We then aggregate individual results to derive a mean cosine distance score for each subreddit so that comparisons can be made across loyal and vagrant user role fluctuations.
In order to observe the variation of community roles over time, we first find the maximum number of clusters present across time periods to be compared by decomposing the 128 dimensional embedding spaces into 2 dimensions using PCA. The Elbow method using Euclidean Kmeans is then applied to determine the number of clusters present. The maximum equal cluster number across two embedding spaces is recorded and k-Nearest Neighbours, where k=1, is applied to compute the Euclidean distance between the closest aligned centroids. The resulting value provides insight into how much the general roles present within a subreddit community have changed over time. Finally, silhouette scores are also computed for each embedding space to determine whether roles evolve to become more or less acutely defined over time. 

\section{Results}

The results of our analysis are depicted in Fig.\ref{fig:results}. The first figure, Fig.\ref{fig:results}(a), illustrates the average cosine distances computed for each subreddit mapped from time period T1 to aligned T2. The majority of user cosine distances continue to remain as dissimilar to each other in the second temporal embedding space, time period T2 aligned with time period T3. Although differences can be observed between loyal and vagrant users, such as vagrant users appearing to change roles to a greater extent than loyal users, while loyal users appear to retain the same role over time. It is hard to define this as a general rule that could be applied to all of Reddit's community without applying our investigation to a larger number of subreddits. However, our preliminary findings suggest that although individual users of Reddit may change role frequently, the universal community level roles remain relatively static in comparison. Fig.\ref{fig:results}(b) depicts how distances between role cluster centroids for both loyal and vagrant subreddits remain small, indicating they are similar to each other.

The static nature of community roles in comparison to user roles is further examined by visualising the PCA projections and by calculating the average Silhouette Score for each subreddit. Visual comparisons of one loyal subreddit (Fig.\ref{fig:vis}(a)(b)), r/Dodgers, and one vagrant subreddit similar in size (Fig.\ref{fig:vis}(c)(d)), r/FoodPorn, depict striking differences in the definition of clusters. The loyal subreddit role clusters are more dispersed in comparison to it's vagrant counterpart where clusters are spread and tightly compact. The average Silhouette scores, Fig.\ref{fig:results}, indicate that it's not an isolated scenario. However, the differences are small, and again, further subreddits will need to be examined before we can say definitively that such differences are significant.


\section{Conclusions and Future Work}

In this paper, we have analysed 29 subreddits classed as either ``loyal'' or ``vagrant'' using a methodology inspired by the study of diachronic word embeddings in the field of natural language processing. Specifically, we applied the role embedding algorithm, \textit{struc2vec} to three consecutive temporal windows of user networks and then aligned the resulting embedding spaces using orthogonal Procrustes. We found that in certain community role cases, orthogonal Procrustes was not enough to align spaces entirely if the subset of overlapping users were not evenly distributed across the embedding space. We then applied a secondary alignment to the principal components to account for it. Overall, our findings suggest that while participant roles fluctuate a lot, the ubiquitous community roles present are a lot more static. However, further analysis is required and we hope to extend the current work to explore subreddits such as AskReddits, Debate Reddits, Questions Reddits, where roles are generally quite distinguished to allow for further comparisons to be made. We also hope to incorporate more measures to further assess temporal changes of roles.

\begin{acks}
  The authors would like to thank Leonardo F.R. Ribeiro for having a public implementation of \textit{struc2vec} available and William L. Hamilton also hosting his Reddit loyalty network data publically. The work is supported by the Science Foundation Ireland under Grant Number SFI/12/RC/2289.

\end{acks}

\bibliographystyle{ACM-Reference-Format}
\bibliography{library}

\end{document}